\documentclass[graphicx,psfig]{llncs}

\usepackage{times}
\usepackage{mathrsfs}
\usepackage{algorithmic,algorithm,float}
\usepackage{amsmath,amssymb}
\usepackage{latexsym}
\usepackage{xspace}
\usepackage{subfigure}
\usepackage{graphicx,color}
\usepackage{paralist}
\usepackage{proof}
\setboolean{ALC@noend}{true}

\newcommand{\powerset}[1]{\mathcal{P}(#1)}

\newcommand{\xacu}{{\sf XAcU}}
\newcommand{\xacuannot}{\ensuremath{\mbox{\xacu}^{annot}}}

\newcommand{\elem}[1]{\ensuremath{#1}}
\newcommand{\EMPTY}{\ensuremath{\mathsf{EMPTY}}}
\newcommand{\str}{\ensuremath{\mathsf{str}}}

\newcommand{\parent}{\ensuremath{\mathsf{parent}}}

\newcommand{\UAT}{{\em UAT}\xspace}
\newcommand{\UATs}{{\em UATs}\xspace}

\newcommand{\Labels}{{\cal L}}

\newcommand{\kinto}{\ensuremath{\mathsf{into}}\xspace}
\newcommand{\kfirst}{\ensuremath{\mathsf{first}}\xspace}
\newcommand{\klast}{\ensuremath{\mathsf{last}}\xspace}
\newcommand{\kbefore}{\ensuremath{\mathsf{before}}\xspace}
\newcommand{\kafter}{\ensuremath{\mathsf{after}}\xspace}
\newcommand{\rdelete}{\ensuremath{\mathsf{delete}}\xspace}
\newcommand{\rinsert}{\ensuremath{\mathsf{insert}}\xspace}
\newcommand{\rreplace}{\ensuremath{\mathsf{replace}}\xspace}
\newcommand{\kas}{\ensuremath{\mathsf{as}}\xspace}

\renewcommand{\paragraph}[1]{\noindent{\bf #1}}

\newcommand{\semantics}[2]{\ensuremath{[\![ #1 ]\!](#2)}}

\renewcommand{\ni}{\noindent}

\newcommand{\ie}{\emph{i.e.,}\xspace}
\newcommand{\eg}{\emph{e.g.,}\xspace}
\newcommand{\wrt}{\emph{w.r.t.}\xspace}

\newcommand{\A}{{\cal A}}
\newcommand{\F}{{\cal F}}
\newcommand{\G}{{\cal G}}
\newcommand{\V}{{\cal V}}
\newcommand{\E}{{\cal E}}

\newcommand{\Se}{{\cal S}}
\newcommand{\U}{{\cal U}}
\newcommand{\C}{{\cal C}}
\newcommand{\I}{{\cal I}}
\newcommand{\sce}{\ensuremath{{\cal E}_{sc}}}
\newcommand{\ra}{\rightarrow}
\newcommand{\la}{\leftarrow}

\newcommand{\J}{{\cal J}}

\usepackage{enumitem}
\setlist{topsep=0pt,noitemsep} 

\newcommand{\NP}{{\sc np}\xspace}

\newcommand{\PTIME}{{\sc ptime}\xspace}

\newcommand{\eop}{\hspace*{\fill}\mbox{$\Box$}}     

\newcommand{\kw}[1]{{\ensuremath {\mathsf{#1}}}\xspace}
\newcommand{\valid}{\kw{valid}}
\newcommand{\isvalid}{\kw{valid\_in}}

\newcommand{\matches}{\ensuremath{\kw{matches}_{\mathit{t}}}}

\renewcommand{\bar}[1]{\overline{#1}}

\newcommand{\algcom}[1]{{\em /* #1 */}}

\renewcommand{\bar}[1]{\overline{#1}}

\newcommand{\sdg}[1]{\ensuremath{G_{#1}}}
\newcommand{\dg}[1]{\ensuremath{#1}}

\newcommand{\mg}[1]{\ensuremath{MG_{#1}}}

\newcommand{\sqleq}{\sqsubseteq}

\newcommand{\Allow}{\mathcal{A}}
\newcommand{\Deny}{\mathcal{D}}
\newcommand{\lpce}[1]{#1^{\dagger}}

\newcommand{\sm}{\setminus}

\newenvironment{proofOf}[1]{\vspace{3mm}\ni{\bf \em Proof of #1.}}{\eop}

\newcommand{\MNJ}{{\mathfrak J}}

\newcommand{\mathquote}[1]{\text{``$#1$''}}

\newcommand{\naive}{\mathbf{ReplaceNaive}}

\begin{document}

\title{Repairing Inconsistent XML Write-Access Control Policies}

\author{Loreto Bravo \and James Cheney \and Irini Fundulaki \vspace{-3mm}}

\institute{School of Informatics,
University of Edinburgh, UK\\
\{lbravo, jcheney, efountou\}@inf.ed.ac.uk}

\maketitle

\begin{abstract}
  XML access control policies involving updates may contain security
  flaws, here called \emph{inconsistencies}, in which a forbidden
  operation may be simulated by performing a sequence of allowed
  operations.  This paper investigates the problem of deciding whether
  a policy is consistent, and if not, how its inconsistencies can be
  repaired. We consider policies expressed in terms of annotated DTDs
  defining which operations are allowed or denied for the XML trees
  that are instances of the DTD. We show that consistency is decidable
  in \PTIME for such policies and that consistent partial policies can
  be extended to unique ``least-privilege'' consistent total policies.
  We also consider repair problems based on deleting privileges to
  restore consistency, show that finding minimal repairs is
  \NP-complete, and give heuristics for finding repairs.
\end{abstract}

\pagestyle{plain}

\section{Introduction}
\label{sec:intro}

Discretionary access control policies for database systems can be
specified in a number of different ways, for example by storing access
control lists as annotations on the data itself (as in most file
systems), or using rules which can be applied to decide whether to
grant access to protected resources.  In relational databases,
high-level policies that employ rules, roles, and other abstractions
tend to be much easier to understand and maintain than access control
list-based policies; also, they can be implemented efficiently using
static techniques, and can be analyzed off-line for security
vulnerabilities~\cite{1146253}.

Rule-based, fine-grained access control techniques for XML data have
been considered extensively for {\em read-only
  queries}~\cite{fan04:_secur_xml_query_secur_views,1063994,990046,fundulaki07:_formal_xml_acces_contr_for_updat_operat,545190,1178621,505590}.
However, the problem of controlling \emph{write access} is
relatively new and has not received much attention. Authors
in~\cite{545190,505590,lim03:_acces_contr_of_xml_docum} studied
enforcement of write-access control policies following
annotation-based approaches.

In this paper, we build upon the schema-based access control model
introduced by Stoica and Farkas~\cite{stfa02}, refined by Fan, Chan,
and Garofalakis~\cite{fan04:_secur_xml_query_secur_views}, and
extended to write-access control by Fundulaki and
Maneth~\cite{fundulaki07:_formal_xml_acces_contr_for_updat_operat}.
We investigate the problem of checking for, and repairing, a
particular class of vulnerabilities in XML write-access control
policies.  An access control policy specifies which actions to allow
a user to perform based on the syntax of the atomic update, not its
actual behavior.  Thus, it is possible that a single-step action
which is explicitly forbidden by the policy can nevertheless be
simulated by one or more allowed actions. This is what we mean by an
\emph{inconsistency}; a consistent policy is one in which such
inconsistencies are not possible. We believe inconsistencies are an
interesting class of policy-level security vulnerabilities since
such policies allow users to circumvent the intended effect of the
policy.  The purpose of this paper is to define consistency,
understand how to determine whether a policy is consistent, and show
how to automatically identify possible repairs for inconsistent
policies.

\begin{figure}[!tb]
\vspace{-2ex}
\begin{center}
\fbox{
\includegraphics[scale=0.4]{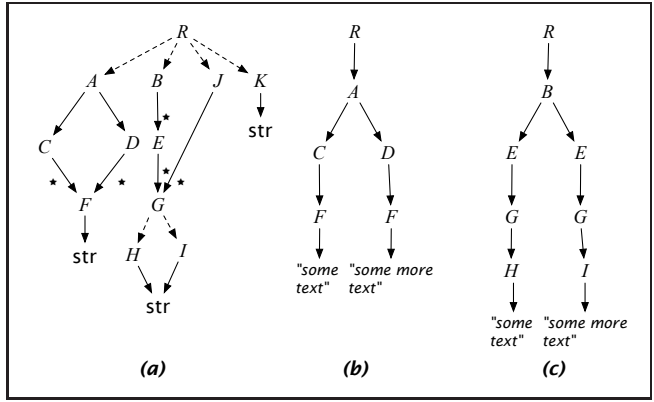}}
\caption{DTD graph (a) and XML documents conforming to the DTD (b, c)}
\end{center}
\label{fig:example}\vspace{-3ex}
\end{figure}

\noindent
{\bf Motivating Example:}  We introduce here an example
and refer to it throughout the paper.
Consider the XML DTD represented as a graph in
Fig.~\ref{fig:example}(a).
A document conforming to this DTD has as
root an $\elem{R}$-element with a single child element that can either be
an $\elem{A}$, $\elem{B}$, $\elem{J}$
or $\elem{K}$-element (indicated with dashed edges); similarly
for $\elem{G}$. An $\elem{A}$-element has one $\elem{C}$ and one $\elem{D}$
children elements. A $\elem{B}$-element can have zero or more
$\elem{E}$ children elements (indicated with $*$-labeled edges);
similarly, $\elem{E}$ and $\elem{J}$ elements can have zero
or more $\elem{G}$ children elements. Finally, $\elem{F}$,
$\elem{H}$, $\elem{I}$ and $\elem{K}$ are text
elements. Fig.~\ref{fig:example}(b) and (c) show two documents that
conform to the DTD.

Suppose that a security policy {\em allows} one to {\em insert} and
{\em delete} $\elem{G}$ elements and {\em forbids} one from
replacing an $\elem{H}$ with an $\elem{I}$ element. It is
straightforward to see that the forbidden operation can be simulated
by first deleting the $\elem{G}$ element with an $\elem{H}$ child
and then inserting a $\elem{G}$ element with an $\elem{I}$ child.
There are different ways of fixing this inconsistency: either {\em (a)}
to allow all operations below element $\elem{G}$ or {\em (b)}
forbid one of the {\em insert} and {\em delete} operations at node
$\elem{G}$.

Now, suppose that the policy {\em allows} one to {\em replace} an
$\elem{A}$-element with a $\elem{B}$-element and this with a
$\elem{J}$-element, but {\em forbids} the replacement of $\elem{A}$
with $\elem{J}$ elements. The latter operation can be easily
simulated by performing a sequence of the allowed operations. As in
the previous case, the repairs that one can propose are {\em (a)} to
allow the forbidden replace operation or {\em (b)} forbid one of the
allowed operations.

\textbf{Our contributions:} In this paper we consider policies that
are defined in terms of {\em non-recursive structured} XML DTDs as
introduced in~\cite{fan04:_secur_xml_query_secur_views} that capture
without loss of generality more general non-recursive DTDs. We first
consider \emph{total} policies in which all allowed or forbidden
privileges are explicitly specified. We define consistency for such
policies and prove the correctness of a straightforward polynomial
time algorithm for consistency checking.  We also consider
\emph{partial} policies in which privileges may be omitted. Such a
policy is consistent if it can be extended to a consistent total
policy; there may be many such extensions, but we identify a
canonical \emph{least-privilege} consistent extension, and show that
this can be found in polynomial time (if it exists).  Finally, given
an inconsistent (partial or total) policy, we consider the problem
of finding a ``repair'', or minimal changes to the policy which
restore consistency.  We consider repairs based on changing
operations from allowed to forbidden, show that finding minimal
repairs is \NP-complete, and provide heuristic repair algorithms
that run in polynomial time.

The rest of this paper is structured as follows: in
Section~\ref{sec:prelims} we provide the definitions for XML DTDs
and trees. Section~\ref{sec:xacu} discusses {\em i)} the atomic
updates and {\em ii)} the access control policies that we are
considering. Consistency is discussed in
Section~\ref{sec:consistent-policies}; Section~\ref{sec:rep}
discusses algorithms for detecting and repairing inconsistent
policies. We conclude in Section~\ref{sec:conclusions}.  Proofs of
theorems and detailed algorithms can be found in the Appendix.


\vspace{-1mm}\section{XML  DTDs and Trees} \label{sec:prelims}

We consider {\em structured} XML DTDs as discussed
in~\cite{fan04:_secur_xml_query_secur_views}.  Although not all DTDs
are syntactically representable in this form, one can (as argued
by~\cite{fan04:_secur_xml_query_secur_views}) represent more general
DTDs by introducing new element types. The DTDs we
consider here are 1-unambiguous as required by the XML
standard~\cite{http:xml}.

\begin{definition}[XML DTD] \em
\label{def:dtd} Let $\Labels$ be the infinite domain of labels.
  A DTD $D$ is represented by $(Ele, Rg, rt)$ where {\em i)} $Ele
  \subseteq \Labels$ is a finite set of {\em element types} {\em ii)} $rt$ is a
  distinguished type in $Ele$ called the {\em root type} and {\em iii)} $Rg$
  defines the element types: that is, for any $A \in Ele$, $Rg(A)$ is
  a regular expression of the form:

\vspace{-3mm}$$  Rg(A)  :=  \str \, \mid \, \epsilon \, \mid \,
B_1,B_2,\ldots, B_n \, \mid \, B_1+B_2+\ldots+B_n\, \mid \, B_1*$$

\vspace{-2mm}\ni where $B_i \in Ele$ are distinct, $\mathquote{,}$,
$\mathquote{+}$ and $\mathquote{*}$ stand for {\em concatenation},
{\em disjunction} and {\em Kleene star} respectively, $\epsilon$ for
the \EMPTY\ element content and \str\ for text values.
\end{definition}
We will refer to $A\rightarrow Rg(A)$ as the {\em production
  rule} for $A$. An element type $B_i$ that appears in the production
rule of an element type $A$ is called the {\em subelement} type of
$A$.  We write $A \leq_D B$ for the transitive, reflexive closure of
the subelement relation.

A DTD can also be represented as a directed acyclic graph that we
call {\em DTD graph}.

\begin{definition}[DTD Graph] \em
  A DTD graph $\sdg{D} = (\V_{\dg{D}}, \E_{\dg{D}}, r_{\dg{D}})$ for a
  DTD $D = (Ele, Rg, rt)$ is a directed acyclic graph (DAG) where {\em i)} $\V_{\dg{D}}$ is the set
  of nodes for the element types in $Ele \cup \{\str\}$, {\em ii)}
$\E_{\dg{D}}=\{(A,B)$ $\mid$ $A,B \in Ele$ and $B$ is a subelement
type of $A\}$ and {\em iii)} $r_{\dg{D}}$ is the distinguished node
$rt$.
\end{definition}

\begin{example}
  The production rules for the DTD graph shown in Fig.~\ref{fig:example}
  are:

\vspace{-4mm}
\begin{small}
\begin{multicols}{4}
\noindent $R \ra A+B+J+K$ \\
\noindent $A\ra C,D$ \\
\noindent $C\ra F*$ \\
\noindent $D\ra F*$ \\
\noindent $B\ra E*$ \\
\noindent $E\ra G*$ \\
\noindent $G\ra H+I$ \\
\noindent $J\ra G*$\\
\noindent $F\ra \str$ \\
\noindent $H\ra \str$ \\
\noindent $I\ra \str$ \\
\noindent $K\ra \str$
\end{multicols}\end{small}
\end{example}

\vspace{-2mm}

\ni We model XML documents as {\em rooted unordered} trees with
labels from $\Labels \cup \{ \str \}$.
\begin{definition}[XML Tree] \em
  \label{def:tree} An unordered XML tree $t$ is an expression of the
  form $t = (N_{t}, E_{t}, \lambda_{t}, r_{t}, v_t)$ where {\em i)}
  $N_{t}$ is the set of nodes {\em ii)} $E_{t} \subset N_{t} \times
  N_{t}$ is the set of edges, {\em iii)} $\lambda_{t}: N_{t}
  \rightarrow \Labels \cup \{ \str \}$ is a labeling function over
  nodes {\em iv)} $r_{t}$ is the root of $t$ and is a distinguished
  node in $N_{t}$ and {\em v)} $v_t$ is a function that assigns a string value
  to nodes labeled with $\str$.
\end{definition}
We denote by $\kw{children}_{t}(n)$, $\kw{parent}_{t}(n)$ and
$\kw{desc}_{t}(n)$,  the children, parent and descendant nodes,
respectively, of a node $n$ in an XML tree $t$. The set
$\kw{desc}^{e}_{t}(n)$ denotes the edges in $E_{t}$ between
descendant nodes of $n$. A node labeled with an element type $A$ in
DTD $D$ is called an {\em instance} of $A$.

We say that an XML tree $t$ $=$ $(N_{t},$ $E_{t},$ $\lambda_{t},$
$r_{t}$, $v_t$$)$  {\em conforms} to a DTD $D = (Ele, Rg,$ $rt)$ at
element type $A$ if {\em i)} $r_{t}$ is labeled with $A$ (i.e.,
$\lambda_{t}(r_{t}) = A$) {\em ii)} each node in $N_{t}$ is labeled
with either an $Ele$ element type $B$ or with $\str$, {\em iii)}
each node in $t$ labeled with an $Ele$ element type $B$ has a list
of children nodes such that their labels are in the language defined
by $Rg(B)$ and {\em iv)} each  node in $t$ labeled with $\str$ has a
string value ($v_t(n)$ is defined) and is a leaf of the tree.  An
XML tree $t$ is a valid instance of the DTD $D$ if $r_{t}$ is
labeled with $rt$.  We write $I_{D}(A)$ for the set of valid
instances of $D$ at element type $A$, and $I_D$ for $I_{D}(rt)$.

\begin{definition}[XML Tree Isomorphism] \em
  We say that an XML tree $t_1$
  is
  isomorphic to an XML tree $t_2$,
  denoted $t_1 \equiv t_2$, iff there exists a bijection $h : N_{t_1}
  \to N_{t_2}$ where: {\em i)} $h(r_{t_1}) = r_{t_2}$ {\em ii)} if
  $(x,y) \in E_{t_1}$ then $(h(x),h(y)) \in E_{t_2}$, {\em iii)}
  $\lambda_{t_1}(x) = \lambda_{t_2}(h(x))$, and {\em iv)} $v_{t_1}(x) =
  v_{t_2}(h(x))$ for every $x$ with $\lambda_{t_1}(x) = \str =
  \lambda_{t_2}(h(x))$.
\end{definition}


\section{XML Access Control Framework}
\label{sec:xacu}

\subsection{Atomic Updates}

Our updates are modeled on the XQuery Update Facility
draft~\cite{http:xqupdate}, which considers $\rdelete$, $\rreplace$
and several $\rinsert$ update operations.  A $\rdelete(n)$ operation
will delete node $n$ and all its descendants. A $\rreplace(n,t)$
operation will replace the subtree with root $n$ by the tree $t$. A
$\rreplace(n,s)$ operation will replace the text value of node $n$ with string
 $s$. There are several types of insert operations, \eg
$\rinsert~\kinto(n,t)$, $\rinsert~\kbefore(n,t)$,
$\rinsert~\kafter(n,t)$, $\rinsert~\kas~\kfirst(n,t)$,
$\rinsert~\kas~\klast(n,t)$. Update $\rinsert~\kinto(n,t)$ inserts
the root of $t$ as a child of $n$ whereas update
$\rinsert~\kas~\kfirst(n,t)$ ($\rinsert~\kas~\klast(n,t)$) inserts
the root of $t$ as a first (resp. last) child of $n$. Update
operations $\rinsert~\kbefore(n,t)$ and $\rinsert~\kafter(n,t)$ insert
the root node of $t$ as a preceding and following sibling of $n$
resp..

Since we only consider unordered XML trees, we deal only with the
operation $\rinsert~\kinto(n,t)$ (for readability purposes, we
are going to write $\rinsert(n,t)$). Thus, in what follows, we will
restrict to four types of update operations: $\rdelete(n)$,
$\rreplace(n,t)$, $\rreplace(n,s)$ and $\rinsert(n,t)$.

More formally, for a tree $t_1$ $=$ $(N_{t_1},$$E_{t_1},$
$\lambda_{t_1}$, $r_{t_1},$ $v_{t_1}$$)$, a node $n$ in $t_1$, a tree
$t_2$ $=$ $(N_{t_2},$ $E_{t_2},$ $\lambda_{t_2},$ $r_{t_2},$
$v_{t_2})$ and a string value $s$,  the result of applying
$\rinsert(n,t_2)$, $\rreplace(n,t_2)$, $\rdelete(n)$ and
$\rreplace(n,s)$ to $t_1$, is a new tree $t=(N_t,E_t,\lambda_t,r_t,v_t)$
defined as shown in Table~\ref{table:UpSem}. We denote by
$\semantics{op}{t}$ the result of applying update operation $op$ on
tree $t$.

\begin{table}[tb]
\vspace{-2ex}
\begin{scriptsize}
\begin{tabular}{|l|l|l|l|l|l|} \cline{2-6}
 \multicolumn{1}{c}{} & \multicolumn{1}{|c|}{$N_{t}$} & \multicolumn{1}{|c|}{$E_{t}$} &
 \multicolumn{1}{|c|}{$\lambda_{t}$} & \multicolumn{1}{|c|}{$r_{t}$} & \multicolumn{1}{|c|}{$v_{t}$}  \\ \hline \hline
$\semantics{\rinsert(n,t_2)}{t_1}$ &
$N_{t_1} \cup N_{t_2}$ &
$E_{t_1} \cup E_{t_2} \cup $ $\{ (n, r_{t_2}) \} $ &
$\lambda_{t_1}(m)$, $m \in N_{t_1}$&
$r_{t_1}$ &
$v_{t_1}(m)$, $m \in N_{t_1}$\\
& &  & $\lambda_{t_2}(m)$,
$m \in N_{t_2}$  & & $v_{t_2}(m)$, $m \in N_{t_2}$\\ \hline
$\semantics{\rreplace(n,t_2)}{t_1}$ &
$N_{t_1} \cup N_{t_2}$&
$E_{t_1} \cup E_{t_2} \cup $&
$\lambda_{t_1}(m)$, &
$r_{t_1}$ &
$v_{t_1}(m)$, $m \in N_{t_1}$\\
& $ \setminus \kw{desc}_{t_1}(n)$ &
 $ \{ (\kw{parent}_{t_1}(n),r_{t_2}) \} \setminus$ &
$m \in (N_{t_1}\setminus\{n\})$ & & $v_{t_2}(m)$, $m \in N_{t_2}$ \\
& & $\kw{desc}^{e}_{t_1}(n)$  & $\lambda_{t_2}(m)$, $m \in N_{t_2}$  & & \\ \hline
$\semantics{\rreplace(n,s)}{t_1}$ &
$N_{t_1}$ &
$E_{t_1}$  &
$\lambda_{t_1}(m)$, $m \in N_{t_1}$ &
$r_{t_1}$ &
$v_{t_1}(m)$, \\
& & & & & $m\in(N_{t_1}\!\! \setminus \!\! \{n\})$ \\
& & & & & $v_{t_1}(n) = s$ \\
\hline
$\semantics{\rdelete(n)}{t_1}$ &
$N_{t_1}\setminus \kw{desc}_{t_1}(n) $ &
$E_{t_1}  \setminus \kw{desc}^{e}_{t_1}(n)$ &
$\lambda_{t_1}(m)$, &
$r_{t_1}$ &
$v_{t_1}(m)$, \\
& & & $m \in (N_{t_1}\!\! \setminus\!\! \kw{desc}_{t_1}(n))$ & & $m \in (N_{t_1}\!\! \setminus\!\! \kw{desc}_{t_1}(n))$ \\ \hline
\end{tabular}
\end{scriptsize}
\caption{Semantics of update operations}
\label{table:UpSem}\vspace{-3ex}
\end{table}
An update operation $\rinsert(n,t_2)$, $\rreplace(n,t_2)$,
$\rreplace(n,s)$ or $\rdelete(n)$ is {\em valid} with respect to tree
$t_1$ provided $n \in N_{t_1}$ and $t_2$, if present, does not
overlap with $t_1$ (that is, $N_{t_1} \cap N_{t_2} = \emptyset$).  We
also consider \emph{update sequences} $op_1;\ldots;op_n$ with the
(standard) semantics $\semantics{op_1;\ldots;op_n}{t_1} =
\semantics{op_n}{\semantics{op_{n-1}}{{\cdots\semantics{op_1}{t_1}}}}$.
A sequence of updates $op_1;\ldots;op_n$ is valid with respect to
$t_0$ if for each $i \in \{1,\ldots,n\}$, $op_{i+1}$ is valid with
respect to $t_i$, where $t_1 = \semantics{op_1}{t_0}$, $t_2 =
\semantics{op_2}{t_1}$, \textit{etc}.  The result of a valid update
(or valid sequence of updates) exists and is unique up to tree
isomorphism.

\subsection{Access Control Framework}

We use the notion of {\em update access type} to specify the access
authorizations in our context. Our update access types are inspired
from the $\xacuannot$ language discussed
in~\cite{fundulaki07:_formal_xml_acces_contr_for_updat_operat}.
Authors followed the idea of {\em security annotations } introduced
in~\cite{fan04:_secur_xml_query_secur_views} to specify the access
authorizations for XML documents in the presence of a DTD.

\begin{definition}[Update Access Types] \em Given a DTD $D$, an {\em
    update access type} (\UAT) defined over $D$ is of the form $(A,\rinsert(B_1))$,
  $(A,\rreplace(B_1,B_2))$, $(A,\rreplace(\str,$ $\str))$ or
  $(A,\rdelete(B_1))$, where $A$ is an element type in $D$, $B_1$
  and   $B_2$ are subelement types of $A$  and $B_1 \neq B_2$.
\end{definition}
Intuitively, an \UAT represents a set of {\em atomic update
operations}. More specifically, for $t$ an instance of DTD $D$,
$op$ an atomic update and  $uat$  an update access type we say that $op$
$matches$ $uat$ on $t$ ($op$ $\matches$ $uat$) if:
\begin{small}
\[\begin{array}{c}
  \infer{\rinsert(n,t')~ \matches~ (A,\rinsert(B))}
  {\lambda_t(n) = A &
    t' \in I_D(B)}
  \quad
  \infer{\rdelete(n)~ \matches~ (A,\rdelete(B))}
  {\lambda_t(n) = B &
    \lambda_t(\parent_t(n))=A}
\smallskip\\
  \infer{\rreplace(n,t')~ \matches~ (A,\rreplace(B, B'))}{\lambda_t(n) = B, t' \in I_D(B'), \lambda_t(parent_t(n))=A, B\neq B'}
\smallskip\\
\infer{\rreplace(n,s)~ \matches (A, \rreplace(\str,\str))}{\lambda_t(n) = \str, \lambda_t(\parent_t(n)) = A\ }
\end{array}\]
\end{small}

\vspace{-2mm} It is trivial to translate our update access types to
$\xacuannot$ security annotations.  In this work we assume that the
evaluation of an update operation on a tree that conforms to a DTD
$D$ results in a {\em tree that conforms to $D$}. It is clear then
that each update access type only makes sense for specific element
types. For our example DTD, the update access type $(A,
\rdelete(C))$ is not meaningful because allowing the deletion of a
$C$-element would result in an XML document that does not conform to
the DTD, and therefore, the update will be rejected. Similar for
$(R, \rdelete(A))$ or $(R, \rinsert(A))$. But, $(B, \rdelete(E))$
and $(B, \rinsert(E))$ are relevant for this specific DTD. The
relation $uat ~\isvalid~ D$, which indicates that an update access
type $uat$ is valid for the DTD $D$, is defined as follows:
\begin{small}
\[\begin{array}{c}
  \infer{(A, \rinsert(B_1)) ~\isvalid~ D}{Rg(A):=B_1^{*}}
  \quad
  \infer{(A, \rdelete(B_1)) ~\isvalid~ D}{ Rg(A):=B_1*}
  \smallskip\\
  \infer{(A,\rreplace(\str,\str)) ~\isvalid~ D}{Rg(A):=\str}
  \quad
  \infer{(A,\rreplace(B_i,B_j)) ~\isvalid~ D}{Rg(A):=B_1+ \dots + B_n, i,j \in [1,n] & i \neq j}
\end{array}
\]
\end{small}

\vspace{-2mm} \ni We define the set of valid \UATs for a given DTD
$D$ as $\valid(D) = \{uat \mid uat $ $\isvalid$ $D\}$.  A {\em
security policy} will be defined by a set of {\em allowed} and {\em
forbidden} valid \UATs.

\begin{definition} \em A security policy $P$ defined over a DTD $D$,
  is represented by $(\A, \F)$ where $\A$ is the set of {\em allowed}
  and $\F$ the set of {\em forbidden} update access types defined over
$D$ such that $\A
  \subseteq \kw{valid}(D)$, $\F \subseteq \kw{valid}(D)$ and $\A \cap
  \F= \emptyset$.
A security policy is {\em total} if ${\cal A} \cup {\cal
  F} = \kw{valid}(D)$, otherwise it is {\em partial}.
\end{definition}

\begin{example} \label{ex:policy}Consider the DTD $D$ in
  Fig.~\ref{fig:example} and the total policy $P\!=\!(\A,\F)$ where $\A$
  is:

\begin{small}
\begin{tabular}{llll} 
$(R, \rreplace(A,B))$ & $(R, \rreplace(B, J))$ & $(R, \rreplace(J, K))$ & $(R, \rreplace(K, J))$ \\
$(R, \rreplace(K, B))$ & $(C, \rinsert(F))$ & $(C, \rdelete(F))$ & $(D, \rinsert(F))$ \\
$(D, \rdelete(F))$ & $(F, \rreplace(\str, \str))$ & $(B, \rinsert(E))$ & $(B, \rdelete(E))$ \\
$(E, \rinsert(G))$ & $(E, \rdelete(G))$ & $(G,\rreplace(I, H))$  & $(J, \rinsert(G))~$ \\
$(J, \rdelete(G))$ & $(D, \rinsert(F))$ & $(D, \rdelete(F))$ & $(H, \rreplace(\str, \str))$ \\
$(I,\rreplace(\str, \str))$  & $(K, \rreplace(\str,\str))$ & & \\ 
\end{tabular}
\end{small}

\ni and $\F =\valid(D) \setminus  \A$.  On the other hand,
$P=(\A,\emptyset)$ is a partial policy. \eop
\end{example}
The operations that are allowed by a policy $P = (\A, \F)$ on an XML tree $t$, denoted
by  $\semantics{\A}{t}$, are  the union of the atomic update
operations matching each  \UAT in $\A$. More formally,
$\semantics{\A}{t}$ $=$ $\{op$ $|$ $op$ \matches $uat$ on $t$, and
$uat \in \A\}$. We say an update sequence $op_1;\ldots;op_n$ is
allowed on $t$ provided the sequence is valid on $t$ and $op_1 \in
\semantics{\A}{t}$, $op_2 \in \semantics{\A}{\semantics{op_1}{t}}$,
etc.  \footnote{Note that this
  is \emph{not} the same as $\{op_1,\ldots,op_n\} \subseteq
  \semantics{\A}{t}$.}  Analogously, the forbidden operations are
$\semantics{\F}{t}$ $=$ $\{op$ $|$ $op$ \matches $uat$ on $t$, and
$uat \in \F\}$. If a policy $P$ is {\em total}, its semantics is
given by its allowed updates, i.e. $\semantics{P}{t}$ $=$
$\semantics{\A}{t}$. The semantics of a partial policy is studied in
detail in Section~\ref{sec:partialpolicies}.


\vspace{-1mm}\section{Consistent Policies}
\label{sec:consistent-policies}

\vspace{-1mm}A policy is said to be consistent if it is not possible
to simulate a forbidden update through a sequence of allowed updates.
More formally:

\begin{definition}\label{def:cons}
  A policy $P=(\A,\F)$ defined over $D$ is consistent if for every
  XML tree $t$ that conforms to $D$, there does not exist a sequence
  $op_1;\dots;op_n$ of updates that is allowed on $t$
  and an update $op_0 \in \semantics{\F}{t}$ such that:
$$\semantics{op_1;\dots;op_n}{t}\equiv \semantics{op_0}{t}.$$
\end{definition}
In our framework inconsistencies can be classified as:
insert/delete and replace.

Inconsistencies due to {\em insert/delete} operations arise when the
policy {\em allows} one to insert {\em and} delete nodes of element
type $A$ whilst {\em forbidding} some operation in some descendant
element type of the node. In this case, the forbidden operation can be
simulated by first deleting an $A$-element and then inserting a new
$A$-element after having done the necessary modifications.

There are two kinds of inconsistencies created by {\em replace}
operations on a production rule $A \to B_1+\cdots+B_n$ of a DTD.
First, if we are allowed to replace $B_i$ by $B_j$ and $B_j$ by $B_k$
but not $B_i$ by $B_k$, then one can simulate the latter operation by a
sequence of the first two. Second, consider that we are allowed to
replace some element type $B_i$ with an element type $B_j$ and vice
versa. If some operation in the subtree of \emph{ either} $B_i$ or
$B_j$ is forbidden, then it is evident that one can simulate the
forbidden operation by a sequence of allowed operations, leading to an
inconsistency.

We
say that \emph{nothing is forbidden below $A$} in a policy
$P\!=\!(\A,\F)$ defined over $D$ if for every $B_i$ s.t. $A \leq_D
B_i$, $(B_i,op) \not \in \F$ for every $(B_i,op) \in \valid(D)$.  If
$A \rightarrow B_1+\ldots+B_n$, then we define the {\em replace
graph} $\G_A = (\V_A,E_A)$ where {\em i)} $\V_A$ is the set of nodes
for $B_1,B_2,\ldots B_n$ and {\em ii)} $(B_i,B_j) \in \V_A$ if there
exists $(A,\rreplace(B_i,B_j)) \in \A$. Also, the set of {\em
forbidden edges} of $A$, is $\E_A=\{(B_i,B_j) \mid
(A,\rreplace(B_i,B_j)) \in \F\}$. We say that a graph $\G\!=\!(\V,\E)$ is \emph{transitive} if
$(x,y),(y,z) \in \E$ then $(x,z) \in \E$.
We write $\G_A^{+}$ for the
transitive graph of $\G_A$. The following theorem characterizes
policy consistency:

\begin{theorem}
\label{thm:TPconsis} A policy $P=(\A,\F)$ defined over DTD $D$ is
consistent if and only if for every production rule:
\begin{enumerate}
\item $A \rightarrow B*$ in $D$, if $(A,\rinsert(B)) \in \A$ and
    $(A,\rdelete(B)) \in \A$, then nothing is forbidden below $B$
\item $A \ra B_1+\cdots+B_n$ in $D$, for every edge $(B_i,B_j)$
    in $\G_A^+$, $(B_i,B_j) \not \in \F_A$, and
  \item $A \ra B_1+\cdots+B_n$ in $D$, if for every $i \in [1,\ldots
    n]$, if $B_i$ is contained in a cycle in $\G_A$
    then nothing is forbidden below
    $B_i$.
\end{enumerate}
\end{theorem}
\begin{proof}[Sketch]
  The forward direction is straightforward, since if any of the rules
  are violated an inconsistency can be found, as sketched above.  For
  the reverse direction, we first need to reduce allowed update
  sequences to certain (allowed) normal forms that are easier to
  analyze, then the reasoning proceeds by cases.  A full proof is
  given in Appendix \ref{app:proof}.\eop
\end{proof}
In the case of total policies, condition 2 in Theorem
\ref{thm:TPconsis} amounts to requiring that the replace graph $\G_A$
is transitive (i.e., $\G_A = \G_A^{+}$)

\begin{example}\label{ex:incPolicy} (example~\ref{ex:policy}
continued) The total policy $P$ is inconsistent because:
\begin{itemize}
    \item  $(E,\rinsert(G))$ and
        $(E,\rdelete(G))$
         are in $\A$, but $(G,\rreplace(H,I))
        \in \F$ (condition 1, Theorem~\ref{thm:TPconsis}),
        \item $(R,\rreplace(A,J))$, $(R,\rreplace(A,K))$ and
            $(R,\rreplace(B,K))$ are in $\F$ (condition 2,
            Theorem~\ref{thm:TPconsis}), and
    \item There are cycles in $\G_R$ involving both $B$ and $J$,
        but below both of them there is a forbidden {\em UAT}, namely
        $(G,\rreplace(H,I))$ (condition 3, Theorem~\ref{thm:TPconsis})
\end{itemize}
\end{example}
It is easy to see that we can check whether properties 1, 2, and 3
hold for a policy using standard graph algorithms:
\begin{proposition} \label{prop:consis}
    The problem of deciding policy consistency is in \PTIME.
\end{proposition}

\begin{remark}
  We wish to emphasize that consistency is highly sensitive to the
  design of policies and update types. For example, we have
  consciously chosen to \emph{omit} an update type $(A,\rreplace(B_i,$
  $B_i))$ for an element type in the DTD whose production rule is
  either of the form $B*$ or $B_1+\ldots +B_n$. Consider the case of a
  conference management system where a \elem{paper} element has a
  \elem{decision} and a \elem{title} subelement. Suppose that the
  policy allows the author of the paper to {\em replace} a
  \elem{paper} with another \elem{paper} element, but forbids to
  change the value of the \elem{decision} subelement. This policy is
  inconsistent since by replacing a \elem{paper} element by another
  with a different \elem{decision} subelement we are able to perform a
  forbidden update. In fact, the
  $\rreplace(\elem{paper},\elem{paper})$ can simulate any other update
  type applying below a $\elem{paper}$ element. Thus, if the policy forbids
  replacement of \elem{paper} nodes, then it would be inconsistent to
  allow any other operation on \elem{decision} and \elem{title}.
  Because of this problem, we argue that update types
  $\rreplace(B_i,B_i)$ should not be used in policies.  Instead, more
  specific privileges should be assigned individually, \eg by allowing
  replacement of the text values of \elem{title} or \elem{decision}.
\end{remark}

\subsection{Partial Policies} \label{sec:partialpolicies}

\emph{Partial policies} may be smaller and easier to maintain than
total policies, but are ambiguous because some permissions are left
unspecified.  An access control mechanism must either allow or deny a
request.  One solution to this problem (in accordance with the
\emph{principle of least privilege}) might be to deny access to the
unspecified operations.  However, there is no guarantee that the
resulting total policy is \emph{consistent}.  Indeed, it is not
obvious that a partial policy (even if consistent) has \emph{any}
consistent total extension.  We will now show how to find consistent
extensions, if they exist, and in particular how to find a
``least-privilege'' consistent extension; these turn out to be unique
when they exist so seem to be a natural choice for defining the
meaning of a partial policy.

For convenience, we write $\A_P$ and $\F_P$ for the allowed and
forbidden sets of a policy $P$; i.e., $P = (\A_P,\F_P)$.  We introduce
an \emph{information ordering} $P\sqsubseteq Q$, defined as $\A_P
\subseteq \A_Q$ and $\F_P \subseteq \F_Q$; that is, $Q$ is ``more
defined'' than $P$.  In this case, we say that $Q$ extends $P$.  We
say that a partial policy $P$ is \emph{quasiconsistent} if it has a
consistent total extension.  For example, a partial policy on the DTD
of Figure~\ref{fig:example} which allows $(B,\rinsert(E))$,
$(B,\rdelete(E))$, and denies $(H,\rreplace(\str,\str))$ is not
quasiconsistent, because any consistent extension of the policy has to
allow $(H,\rreplace(\str,\str))$.

We also introduce a \emph{privilege ordering} on total policies $P
\leq Q$, defined as $\A_P \subseteq \A_Q$; that is, $Q$ allows every
operation that is allowed in $P$.  This ordering has unique greatest
lower bounds $P \wedge Q$ defined as $(\A_P \cap \A_Q,\F_P \cup
\F_Q)$.  We now show that every quasiconsistent policy has a
\emph{least-privilege} consistent extension $\lpce{P}$; that is,
$\lpce{P}$ is consistent and $\lpce{P} \leq Q$ whenever $Q$ is a
consistent extension of $P$.

\begin{lemma}\label{lem:conj-ext}
  If $P_1,P_2$ are consistent total extensions of $P_0$ then $P_1
  \wedge P_2$ is also a consistent extension of $P_0$.
\end{lemma}
\begin{proof}
  It is easy to see that if $P_1,P_2$ extend $P_0$ then $P_1 \wedge
  P_2$ extends $P_0$. Suppose $P_1 \wedge P_2$ is inconsistent.  Then
  there exists an XML tree $t$, an atomic operation $op_0 \in
  \semantics{\F_{P_1\wedge P_2}}{t}$, a sequence $\bar{op} $ allowed on
  $t$ by $P_1\wedge P_2$, such that $\semantics{op_0}{t} =
  \semantics{\bar{op}}{t}$.  Now $\A_{P_1 \wedge P_2} = \A_{P_1}\cap
  \A_{P_2}$, so $op_0$ must be forbidden by either $P_1$ or $P_2$.  On
  the other hand, $\bar{op}$ must be allowed by \emph{both} $P_1$ and
  $P_2$, so $t,op_0,\bar{op}$ forms a counterexample to the
  consistency of $P_1$ (or symmetrically $P_2$). \eop
\end{proof}

\begin{proposition}
  Each quasiconsistent policy $P$ has a unique $\leq$-least consistent
  total extension $\lpce{P}$.
\end{proposition}
\begin{proof}
  Since $P$ is quasiconsistent, the set $S = \{Q \mid P \sqleq Q, \text{$Q$
    consistent}\}$ is finite, nonempty, and closed under $\wedge$, so
  has a $\leq$-least element $\lpce{P} = \bigwedge S$. \eop
\end{proof}
Finally, we show how to find the least-privilege consistent extension,
or determine that none exists (and hence that the partial policy is
not quasiconsistent).  Define the operator $T: \powerset{\valid(D)} \to
\powerset{\valid(D)}$ as:

\vspace{-6mm}
\begin{small}
\begin{eqnarray*}
  T(S) &=& S \cup \{(C,uat) \mid B \leq_D C, Rg_D(A) = B^*, \{(A,\rinsert(B)), (A,\rdelete(B))\} \subseteq S \}\\
  &&\cup\{(C,uat) \mid B_i \leq_D C, Rg_D(A) = B_1+\ldots+B_n, (B_i,B_i) \in \G_A^+(S) \}\\
  && \cup\{(A,\rreplace(B_i,B_k)) \mid Rg_D(A) = B_1+\ldots+B_n, (B_i,B_k) \in \G_A^+(S)\}
\end{eqnarray*}
\end{small}
\begin{lemma}\label{lem:t-simulation}
  If $uat \in T(S)$ then any operation $op_0$ matching $uat$
  on $t$ can be simulated using a sequence of operations $\bar{op}$
  allowed on $t$ by $S$ (that is, such that $\semantics{op_0}{t} =
  \semantics{\bar{op}}{t}$).
\end{lemma}
\begin{theorem}
Let $P$ be a partial policy.  The following are equivalent: (1) $P$
is quasiconsistent, (2) $P$ is consistent (3) $T(\A_P) \cap \F_P =
\emptyset$.
\end{theorem}
\begin{proof}
  To show (1) implies (2), if $P'$ is a consistent extension of $P$,
  then any inconsistency in $P$ would be an inconsistency in $P'$, so
  $P$ must be consistent.  To show (2) implies (3), we prove the
  contrapositive.  If $T(\A_P) \cap \F_P\neq \emptyset$ then choose $uat
  \in T(\A_P) \cap \F_P$.  Choose an arbitrary tree $t$ and atomic
  update $op$ satisfying $op_0 \in \semantics{uat}{t}$.  By
  Lemma~\ref{lem:t-simulation}, there exists a sequence $\bar{op}$
  allowed by $\A_P$ on $t$ with $\semantics{\bar{op}}{t} =
  \semantics{op_0}{t}$.  Hence, policy $P$ is inconsistent.  Finally, to show
  that (3) implies (1), note that $(T(\A_P),$ $\valid(D)\sm T(\A_P))$
  extends $P$ and is consistent provided $T(\A_P) \cap \F_P =
  \emptyset$.
\end{proof}
Indeed, for a (quasi-)consistent $P$, the least-privilege consistent
extension of $P$ is simply $\lpce{P} = (T(\A_P),\valid(D)\sm T(\A_P))$
(proof omitted).  Hence, we can decide whether a partial policy is
(quasi-)consistent and if so find $\lpce{P}$ in \PTIME.


\section{Repairs}\label{sec:rep}

If a policy is inconsistent, we would like to suggest
possible minimal ways of modifying it in order to restore
consistency. In other words, we would like to find {\em repairs}
that are as close as possible to the inconsistent policy.

There are several ways of defining these repairs.  We might want to
repair by changing the permissions of certain operations from allow to
forbidden and vice versa; or we might give preference to some type of
changes over others. Also, we can measure the minimality of the
repairs as a minimal number of changes or a minimal set of changes under
set inclusion.

Due to space restrictions, in this paper we will focus on finding
repairs that transform \UATs from {\em allowed} to {\em forbidden}
and that minimize the number of changes. We believe that such
repairs are a useful special case, since the repairs are guaranteed
to be more restrictive than the original policy.

\begin{definition}\em
A policy $P'=(\A',\F')$ is a {\em repair} of a policy
$P=(\A,\F )$ defined over a DTD $D$
  iff: i) $P'$ is a policy defined over $D$, ii) $P'$ is
  consistent, and iii) $P' \leq P$.

  A repair is {\em total} if $\F'=\valid(D) \setminus \A$ and {\em
    partial} otherwise. Furthermore a repair $P'=(\A',\F')$ of
  $P(\A,\F )$ is a {\em minimal-total-repair} if there is no total
  repair $P''=(\A'', \F'')$ such that $|\A'|$ $<$ $ |\A''|$ and a {\em
    minimal-partial-repair} if $\F'=\F$ and there is no partial repair
  $P''=(\A'',\F)$ such that $|\A'|$ $<$ $ |\A''|$.
\end{definition}
Given a policy $P=(\A,\F)$ and an integer $k$, the total-repair (partial-repair) problem consists in determining if there exists a total-repair (partial-repair) $P'=(\A',\F')$ of policy $P$ such that $|\A \sm \A'| < k$. This problem can be shown to be \NP-hard by reduction from the edge-deletion transitive-digraph problem \cite{Yan81}.

\begin{theorem} \label{thm:RepProblem}
 The total-repair and partial-repair problem is \NP-complete.
\end{theorem}
If the DTD has no production rules of the type $A \ra B_1 + \dots + B_n$, then the total-repair problem is in {\sc ptime}.

\subsection{Repair Algorithm}

In this section we discuss a repair algorithm that finds a minimal
repair of a total or partial policy. All the algorithms can be found
in Appendix \ref{app:alg}.

The algorithm to compute a minimal repair of a policy relies in the
independence between inconsistencies \wrt insert/delete
(Theorem~\ref{thm:TPconsis}, condition 1) and replace
(Theorem~\ref{thm:TPconsis}, conditions 2 and 3) operations. In
fact, a local repair of an inconsistency \wrt insert/delete
operations will never solve nor create an inconsistency with respect
to a replace operation and vice-versa. We will separately describe
the algorithm for repairing the insert/delete inconsistencies and
then the algorithm for the replace ones.

Both algorithms make use of the {\em marked DTD graph} $\mg{D}=
(\sdg{D}, \mu, \chi)$ where $\mu$ is a function from nodes in
$\V_{\dg{D}}$ to $\{\mathquote{+},\mathquote{-} \}$ and $\chi$ is a
partial function from $\V_{\dg{D}}$ to $\{ \perp \}$.  In a marked
graph for a DTD $D$ and a policy $P = (\A, \F)$ {\em i)} each node
in the graph is either marked with $\mathquote{+}$ (i.e., nothing is
forbidden below the node) or with a $\mathquote{-}$ (i.e., there
exists at least one update access type that is forbidden below the
node). If, for nodes $A$ and $B$ in the DTD, {\em both} $(A,
\rinsert(B))$ and $(A, \rdelete(B))$ are in ${\cal A}$ and
$\mu(A)=\mathquote{-}$, then $\chi(A) = \mathquote{\perp}$. A marked
graph is obtained from algorithm $\mathbf{markGraph}$  which takes
as input a DTD graph and a policy $P$ and traverses the DTD graph
starting from the nodes with out-degree 0 and  marks the nodes and
edges as discussed above.

\begin{figure}[tb]
\vspace{-4ex}
\begin{center}
\fbox{
\includegraphics[scale=0.42]{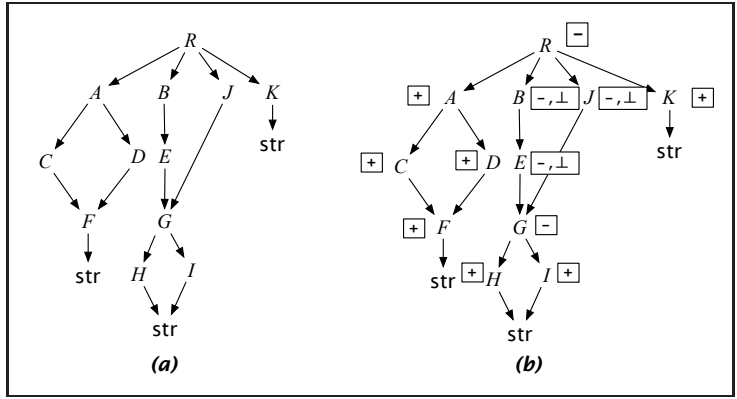}}
\end{center}
\vspace{-1ex}
\caption{DTD Graph (a) and Marked DTD Graph (b) for the DTD in Fig.~\ref{fig:example}}
\label{fig:markgraph}
\vspace{-2ex}\end{figure}

\begin{example} Consider the graph for DTD $D$ in Fig.~\ref{fig:markgraph}(a)  and
policy $P=(\A,\F)$, with $\A$ defined in Example~\ref{ex:policy}.
The result of applying $\mathbf{markGraph}$ to this DTD and policy is
shown in Fig.~\ref{fig:markgraph}(b). Notice that nodes $B$, $E$ and
$J$ are  marked with both a $\mathquote{-}$ and $\mathquote{\perp}$ since {\em i)}
update access type $(G,
\rreplace(H,I))$ is in ${\cal F}$ and {\em ii)}
all insert and delete update access types for $B$, $E$ and $J$
are in ${\cal A}$. For readability purposes
we do not show the multiplicities in the marked DTD graph. \eop
\end{example}

\subsubsection{Repairing Inconsistencies for Insert and Delete
Operations}

Recall that if both the insert and delete operations are allowed at
some element type and there is some operation below this type that
is not allowed, then there is an inconsistency (see
Theorem~\ref{thm:TPconsis}, condition 1). The marked DTD graph
provides exactly this information: a node $A$ is labeled with
$\mathquote{\perp}$ if it is inconsistent w.r.t. {\em insert/delete}
operations. For each such node and for the repair strategy that we
have chosen, the inconsistency can be minimally repaired by removing
either $(A,\rinsert(B))$ or $(A,\rdelete(B))$ from $\A$. Algorithm
$\mathbf{InsDelRepair}$ takes as input a DTD graph $\sdg{D}$ and a
security policy $P = (\A, \F)$ and returns a set of \UATs to remove
from $\A$ to restore consistency \wrt insert/delete-inconsistencies.

\begin{example}
Given the marked DTD graph in Fig.~\ref{fig:markgraph}(b), it is easy
to see that the \UATs that must be repaired are associated with nodes
$B$, $J$ and $E$ (all nodes are marked with $\mathquote{\perp}$). The repairs
that can be proposed to the user are to remove from $\A$ one \UAT
from each of the following sets: $\{(B,\rinsert(E)),$
$(B,\rdelete(E))\}$, $\{(E,\rinsert(G)),$ $(E,\rdelete(G))\}$ and
$\{(J,\rinsert(G)),$ $(J,\rdelete(G))\}$. \eop
\end{example}

\vspace{-3mm}\subsubsection{Repairing Inconsistencies for Replace Operations}

\label{sec:repair-replace} There are two types of inconsistencies
related to replace operations (see Theorem~\ref{thm:TPconsis},
conditions 2--3): the first arises when some element $A$ is contained
in some cycle and something is forbidden below it; the second arises
when the replace graph $\G_A$ cannot be extended to a transitive graph
without adding a forbidden edge in $\F$. In what follows we will
refer to these type of inconsistencies as {\em negative-cycle} and
{\em forbidden-transitivity}. By Theorem~\ref{thm:RepProblem}, the
repair problem is \NP-complete, and therefore, unless {\sc p} = \NP,
there is no polynomial time algorithm to compute a minimal repair to
the replace-inconsistencies. Our objective then, is to find an
algorithm that runs in polynomial time and computes a repair that is
not necessarily minimal.

Algorithm $\naive$ traverses the marked graph $\mg{D}$ and at each
node, checks whether its production rule is of the form $A\to
B_1+\ldots +B_n$. If this is the case, it builds the replace graph
for $A$, $\G_A$,  and runs a modified version of the Floyd-Warshall
algorithm \cite{floyd62:_algor_short_path}. The original
Floyd-Warshall algorithm adds an edge $(B,D)$ to the graph if there
is a node $C$ such that $(B,C)$ and $(C,D)$ are in the graph and
$(B,D)$ is not. Our modification consists on deleting either $(B,C)$
or $(C,D)$ if $(B,D) \in \F_A$, \ie if there is
forbidden-transitivity.  In this way, the final graph will satisfy
condition 2 of Theorem \ref{thm:TPconsis}. Also, if there are edges
$(B,C)$ and $(C,B)$ and $\mu(C)=\mathquote{-}$, \ie there is a
negative-cycle, one of the two edges is deleted. Algorithm $\naive$
returns the set of edges to delete from each node to remove
replace-inconsistencies.

\begin{figure}[tb]
\vspace{-3ex}
\begin{center}
\fbox{
\includegraphics[scale=0.4]{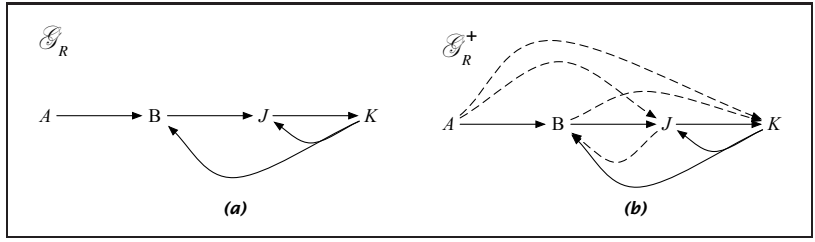}
}
\end{center}\vspace{-2ex}
\caption{Replace $\G_{R}$ (a) and Transitive Replace Graph $\G_{R}^{+}$(b)}
\label{fig:rreplace}
\vspace{-2ex}\end{figure}

\begin{example} The replace graph $\G_G$ has no negative-cycles nor
  forbidden-transitivity, therefore it is not involved in any
  inconsistency. On the other hand, the replace graph $\G_R=(\V,\E)$,
  shown in Fig.~\ref{fig:rreplace}(a) is the source of many inconsistencies. A
  possible execution of $\naive$ (shown in
Fig.~\ref{fig:rreplace-naive} in the Appendix) is: $(A,B),(B,J) \in
\E$ but $(A,J)$ $\in$ $\F$, so $(A,B)$ or $(B,J)$
  should be deleted, say $(A,B)$.  Now, $(B,J),$ $(J,K)$ $\in$ $\E$
  and $(B,K)$ $\in$ $\F$, therefore we delete either $(B,J)$ or
  $(J,K)$, say $(B,J)$. Next, $(K,J),$ $(J,K)$ $\in$ $\E$ and
  $\mu(J)=\mathquote{-}$ in Fig.~\ref{fig:markgraph}(b), therefore
  there is a negative-cycle and either $(K,J)$ or $(J,K)$ has to be
  deleted. If $(K,J)$ is deleted, the resulting graph has no
  forbidden-transitive and nor negative-cycles.  The policy obtained
  by removing $(R,$ $\rreplace(A,B))$, $(R,$ $\rreplace(B,J))$ and
  $(R,$ $\rreplace(J,K))$ from $\A$ has no
  replace-inconsistencies. \eop \end{example}

The $\naive$ algorithm
might remove more than the necessary edges to achieve
consistency: in our example, if we had
removed edge $(B,J)$ at the first step, then we would have resolved
the inconsistencies that involve edges $(A,B)$, $(B,J)$ and $(J,K)$.

An alternative to algorithm $\naive$, that can find a solution
closer to minimal repair, is algorithm $\mathbf{ReplaceSetCover}$,
which also uses a modified version of the Floyd-Warshall algorithm.
In this case, the modification consists in computing the transitive
closure of the replace graph $\G_A$ and labelling each newly
constructed edge $e$ with a set of {\em justifications} $\J$. Each
justification contains sets of edges of $\G_A$ that were used to add
$e$ in $\G_A^+$. Also, if a node is found to be part of a
negative-cycle, it is labelled with the justifications $\J$ of the
edges in each cycle that contains the node. An edge or vertex might
be justified by more than one set of edges. In fact, the number of
justifications an edge or node might have is $O(2^{|\E|})$. To avoid
the exponential number of justifications,
$\mathbf{ReplaceSetCover()}$ assigns at most $\MNJ$ justifications
to each edge or node, where $\MNJ$ is a fixed number. This new
labelled graph is then used to construct an instance of the minimum
set cover problem (MSCP) \cite{Papadimitriou94}. The solution to the
MSCP, can be used to determine the set of edges to remove from
$\G_A$ so that none of the justifications that create
inconsistencies are valid anymore. Because of the upper bound $\MNJ$
on the number of justifications, it might be the case that the graph
still has forbidden-transitive or negative-cycles. Thus, the
justifications have to be computed once more and the set cover run
again until there are no more replace inconsistencies.

\begin{example} \label{ex:labels} For $\MNJ=1$, the first computation of justifications of $\mathbf{ReplaceSetCover}$ results in the graph in  Fig.~\ref{fig:rreplace} (b) with the following justifications:

\vspace{-4mm}
\begin{small}
\begin{multicols}{2}
\ni $\J((A,$ $J))=\{\{(A,$ $B),$ $(B,$ $J)\}\}$

\ni $\J((A,$ $K))=\{\{(A,$ $B),$ $(B,$ $J),$ $(J,$ $K)\}\}$

\ni $\J((B,$ $K))=\{\{(B,$ $J),$ $(J,$ $K)\}\}$

\ni $\J((J,$ $B))=\{\{(J,$ $K),$ $(K,$ $B)\}\}$

\ni $\J(B))=\{\{(B,$ $J),$ $(J,$ $K),$ $(K,$ $B)\}\}$

\ni $\J(J)=\{\{(J,$ $K),$ $(K,$ $J)\}\}$

\end{multicols}\end{small}

\vspace{-3mm}
\ni Justifications for edges represent violations of transitivity. Justification for nodes represent negative-cycles. If we want to remove the inconsistencies, it is enough to
delete one edge from each set in $\J$. \eop
\end{example}
The previous example shows that, for each node $A$, replace-inconsistencies can be repaired by removing at least one edge from each of the justifications  of edges and vertices in $\G_A^+$. It is easy to see that this problem can be reduced to the MSCP. An instance of the MSCP consists of a universe $\mathcal{U}$ and a set $\mathcal{S}$ of subsets of $\mathcal{U}$. A subset $\mathcal{C}$ of $\mathcal{S}$ is a set cover if the union of the elements in it is $\mathcal{U}$. A solution of the MWSCP is a set cover with the minimum number of elements.

The set cover instance associated to $\G_A^+=(\V,\E)$ and the set of forbidden edges $\F_A$, is $\mathit{MSCP}(\G_A^+,\F_A)=(\U,\Se)$ for i) $\U$ $=$ $\{s$ $\mid$ $ s$ $ \in$ $ {\cal J}(e),$ $ e$ $ \in$ $ \F_A \}$ $ \cup$ $ \{s$ $ \mid$ $ s$ $ \in$ $ {\cal J}(V),$ $ V \in$ $ \V\}$, and ii) $\Se$ $=$ $ \bigcup_{e\in \E}$ $ \I(e)$ where $\I(e)$ $=$ $\{s$ $ \mid$ $ s$ $ \in$ $ \U,$ $ e$ $ \in$ $ s\}$. Intuitively, $\U$ contains all the inconsistencies, and the set $\I(e)$ the replace-inconsistencies in which an edge $e$ is involved. Notice that in this instance of the MSCP, the $\U$ is a set of justifications, therefore, $\Se$ is a set of sets of justifications.

\begin{example} The minimum set cover instance,
  $\mathit{MSCP}(\G_R^+,E)=(\U,\Se)$, is such that $\U\!=\!\{\{(A, B),
  (B,J), (J, K)\}, \{(A,B), (B, J)\}, \{(B,J), (J,K)\}, \{(J,K),
  (K,B)\},$ $\{(J,$ $K),$ $ (K,$ $J)\},$ $ \{(K,J),$ $ (J,K)\},$
  $\{(B,J),$ $ (J,K),$ $ (K,B)\}\}$ and $\Se=$ $\{\I((A,B))$,
  $\I((B,J))$, $\I((J,K))$, $\I((K,J))$, $\I((K,B))\}$. The extensions
  of $\I$ are given in Table \ref{tab:SetCover}, where each column
  corresponds to a set $\I$ and each row to an element in $\U$. Values
  1 and 0 in the table represent membership and non-membership
  respectively.  A minimum set cover of $\mathit{MSCP}(\G_R^+)$ is $
  \C=\{\I(B,J),\I(J,K)\}$, since $\I(B,J)$ covers all the elements of
  $\U$ except for the element $\{(A,B),$ $(B,J)\}$, which is covered
  by $\I(J,K)$. Now, using the solution from the set cover, we remove
  edges $(B,J)$ and $(J,K)$ from $\G_R$. If we try to compute the
  justifications once again, it turns out that there are no more
  negative-cycles and that the graph is transitive. Therefore, by
  removing $(R,\rreplace(B,J))$ and $(R,\rreplace(J,$ $K))$ from $\A$,
  there are no replace-inconsistencies in node $R$.  \eop
\end{example}

\begin{table}
\vspace{-4ex}
\begin{center}
\begin{footnotesize}
\begin{tabular}{|c||c|c|c|c|c|}
  \cline{2-6}
  \multicolumn{1}{c||}{} & \multicolumn{5}{|c|}{$\Se$}\\                                \hline
 $\U$   & ~$\I((A,B))$~ & ~$\I((B,J))$~ & ~$\I((J,K))$~ & ~$\I((K,J))$~ & ~$\I((K,B))$~ \\    \hline    \hline
 $ \{(A,B),$ $(B,J),$ $(J,K)\}$ & 1 & 1 & 1 & 0 & 0 \\
  $ \{(A,B),$ $(B,J)\}$ & 1 & 1 & 0 & 0 & 0 \\
 $ \{(B,J),$ $(J,K)\}$ & 0 & 1 & 1 & 0 & 0 \\
      $ \{(J,K),$ $(K,B)\}$ & 0 & 0 & 1 & 0 & 1 \\  
    $ \{(J,K),$ $(K,J)\}$ & 0 & 0 & 1 & 1 & 0 \\
 $ \{(K,J),$ $(J,K)\}$ & 0 & 0 & 1 & 1 & 0 \\
  $ \{(B,J),$ $(J,K),$ $(K,B)\}$ & 0 & 1 & 1 & 0 & 1 \\                    \hline
\end{tabular}\end{footnotesize}\end{center} \caption{Set cover problem \label{tab:SetCover}} \vspace{-4ex}
\end{table}
\ni The set cover problem is MAXSNP-hard \cite{Papadimitriou94}, but
its solution can be approximated in polynomial time using a
greedy-algorithm that can achieve an approximation factor of
$\log(n)$ where $n$ is the size of $\U$~\cite{Chvatal79}. In our
case, $n$ is $O(\MNJ \times |Ele|)$. In the ongoing example, the
approximation algorithm of the set cover will return a cover of size
2. This is better than what was obtained by the $\naive$ algorithm.
In order to decide which one is better, we need to run experiments
to investigate the trade off between efficiency and the size of the
repaired policy.

Algorithm $\mathbf{ReplaceRepair}$ will compute the set of \UATs to
remove from $\A$, by using either $\naive$ (if $\MNJ=0$) or
$\mathbf{ReplaceSetCover}$ (if $\MNJ>0$).

\vspace{-3mm}
\subsubsection{Computation of a Repair}
Algorithm $\mathbf{Repair}$ computes a new consistent policy
$P'=(\A',\F')$ from $P=(\A,\F)$ by removing from $\A$ the union of the
\UATs returned by algorithms $\mathbf{InsDelRepair}$ and
$\mathbf{ReplaceRepair}$. If argument $total$ of algorithm
$\mathbf{Repair}$ is $true,$ then the repair returned by it will be
total. If {\em false}, then a partial policy such that $\F'=\F$ will
be returned.

\begin{theorem}\label{thm:sound}
  Given a total (partial) policy $P$, algorithm $\mathbf{Repair}$
  returns a total (partial) repair of $P$.
\end{theorem}


\section{Conclusion}
\label{sec:conclusions}

Access control policies attempt to constrain the actual operations
users can perform, but are usually enforced in terms of syntactic
representations of the operations. Thus, policies controlling update
access to XML data may forbid certain operations but permit other
operations that have the same effect.  In this paper we have studied
such \emph{inconsistency} vulnerabilities and shown how to check
consistency and repair inconsistent policies.  This is, to our
knowledge, the first investigation of consistency and repairs for XML
update security.  We also considered consistency and repair problems
for partial policies which may be more convenient to write since many
privileges may be left unspecified.

Cautis, Abiteboul and Milo
in~\cite{cautis07:_reason_about_xml_updat_const} discuss XML update
constraints to restrict insert and delete updates, and propose to
detect updates that violate these constraints by measuring the size
of the modification of the database. This approach differs from our
security framework for two reasons: a) we consider in addition to
insert/delete also {\em replace} operations and b) we require that
each operation in the sequence of updates does not violate the
security constraints, whereas in their case, they require that only
the input and output database satisfies them.

Minimal repairs are used in the problem of returning consistent
answers from inconsistent databases~\cite{ABC99}. There, a
consistent answer is defined in terms of all the minimal repairs of a
database. In \cite{BBF+05} the set cover problem was used to find
repairs of databases \wrt denial constraints.

There are a number of possible directions for future work, including
running experiments for the proposed algorithms, studying
consistency for more general security policies specified using XPath
expressions or constraints, investigating the complexity of and
algorithms for other classes of repairs, and considering more
general  DTDs.

\vspace*{2.8ex}

\ni {\bf Acknowledgments:} We would like to thank Sebastian Maneth and
Floris Geerts for insightful discussions and comments.

\vspace*{-.18ex}

\bibliography{paper}
\bibliographystyle{plain}

\newpage
\appendix

\newcommand{\orth}{\mathrel{\bot}}

\section{Proofs} \label{app:proof}

\subsection{Proofs from Section~\ref{sec:consistent-policies}}

In this appendix we outline a detailed proof of correctness for our
characterization of policy consistency (Theorem~\ref{thm:TPconsis}).
The proof is not deep, but requires considering many combinations of
cases.  The main difficulty is in proving that rules 1, 2, and 3 imply
consistency, since this involves showing that for a consistent policy,
there is no way to simulate a single forbidden operation via a
sequence of allowed operations.  The obvious approach by induction on
the length of the allowed sequence does not work because subsequences
of the allowed sequence do not necessarily continue to simulate the
denied operation.

The solution is to establish the existence of an appropriate
\emph{normal form} for update sequences, such that (roughly speaking):
\begin{enumerate}
\item The normal form of an update sequence $\bar{a}$ applied to input
  $t$ is
  \[\rdelete(n_1);\cdots;\rdelete(n_i); \bar{r};
  \rinsert(l_1,v_1),\ldots,\rinsert(l_j,v_j)\]
consisting of a sequence of
  deletes, then replacements, then inserts
\item The replacements $\bar{r}$ can be partitioned into ``chained''
  subsequences $\bar{r_1},\ldots,\bar{r_j}$ that of the form
  $\bar{r_i} = \rreplace(m_i,u_1^i);\rreplace(r_{u_1^i},u_2^i);\cdots$.
\item Each $n_i,m_j, l_k$ is in $t$.
\item No deleted or replaced node ($n_i$ or $m_j$) is an ancestor of
  another of the modified nodes ($n_i,m_j,l_k$)
\item Allowed update sequences have allowed normal forms.
\end{enumerate}
Pictorially, a normalized update sequence can be visualized as a tree
with some of its nodes ``annotated'' with insertion operations
$\rinsert(u)$, deletions $\rdelete$, and replacement sequences
$\rreplace(u_1,\ldots,u_n)$, such that no annotation occurs below a node
with a delete or replace annotation.  Such annotations can be viewed
as instructions for how to construct $\semantics{\bar{a}}{t}$ from
$T$.

Normalized update sequences are much easier to analyze than arbitrary
allowed sequences in the proof of the reverse direction of
Theorem~\ref{thm:TPconsis}.

We introduce some additional helpful notation: write
\begin{eqnarray*}
node(\rdelete(n)) &=&
n\\
 node(\rinsert(n,u)) &=& n
\\
 node(\rreplace(n,u)) &=&n
\end{eqnarray*}
for the ``principal'' node of an operation; write $\leq_t$ for the
ancestor-descendant ordering on $t$ (that is, $E^*$); write
$\orth_{t}$ for the relation $\{(n,m) \in N_t \times N_t \mid n \not\leq_t m
\text{ and } m \not\leq_t n\}$ (that is, $n \orth_t m$ means $n$ and $m$ are
$\leq_t$-incomparable).

\begin{proposition}
  Let $P$ be a security policy and $\bar{a}$  an allowed update
  sequence mapping $t$ to $t'$.  Then there is an equivalent allowed
  update sequence $\bar{a'}$ that is in normal form.
\end{proposition}
\begin{proof}
  We first note that the laws in Figures~\ref{fig:insert-laws},~\ref{fig:delete-laws}, and~\ref{fig:replace-laws} are valid for
  rewriting update sequences relative to a given input tree $t$.  We
  write $\bar{op} \equiv \bar{op'}$ to indicate that the (partial)
  functions $\semantics{\bar{op}}{-}$ and $\semantics{\bar{op'}}{-}$
  are equal; that is, for any tree $t$, $op$ is valid on $t$ if and
  only of $op'$ is valid on $t$, and if both are valid, then
  $\semantics{\bar{op}}{t} = \semantics{\bar{op'}}{t}$.
\begin{figure}
\[\begin{array}{rcl}
\rinsert(n,u);\rinsert(m,v) &\equiv& \left\{
\begin{array}{ll}
\rinsert(n,\semantics{\rinsert(m,v)}{u}) & \text{if $m \in N_u$}\\
\rinsert(m,v);\rinsert(n,u) & \text{if $m \not\in N_u$}
\end{array}
\right.\smallskip\\
\rinsert(n,u);\rreplace(m,v) &\equiv& \left\{
\begin{array}{ll}
\rreplace(m,v) & \text{if $n \in N_t$, $m \leq_t n$}\\
\rreplace(m,v);\rinsert(n,u) & \text{if $n \in N_t$, $m \not\leq_t n$}\\
\rinsert(n,v) & \text{if $m=r_u$}\\
\rinsert(n,\semantics{\rreplace(m,v)}{u}) & \text{if $m \in N_u-\{r_u\}$}
\end{array}
\right.\smallskip\\
\rinsert(n,u);\rdelete(m) &\equiv& \left\{
\begin{array}{ll}
\rdelete(m) & \text{if $m \leq_t n$}\\
\rdelete(m);\rinsert(n,u) & \text{if $m \in N_t$, $m \not\leq_t n$}\\
\epsilon & \text{if $m=r_u$}\\
\rinsert(n,\semantics{\rdelete(m)}{u}) & \text{if $m \in N_u-\{r_u\}$}
\end{array}
\right.
\end{array}\]
\caption{Moving inserts forward}\label{fig:insert-laws}
\[\begin{array}{rcl}
\rreplace(n,u);\rdelete(m) &\equiv& \left\{
\begin{array}{ll}
\rdelete(m) & \text{if $m <_t n$}\\
\rdelete(m);\rreplace(n,u) &\text{if $m \in N_t$, $m \not\leq_t n$}\\
\rdelete(n) &\text{if $m = r_u$}\\
\rreplace(n,\semantics{\rdelete(m)}{u}) &\text{if $m \in N_u-\{r_u\}$}
\end{array}
\right.\smallskip\\
\rdelete(n);\rdelete(m) &\equiv& \left\{
\begin{array}{ll}
\rdelete(m) &\text{if $m \leq_t n$}\\
\rdelete(m);\rdelete(n) & \text{if $m \not\leq_t n$}
\end{array}
\right.
\end{array}
\]
\caption{Moving deletes backward}\label{fig:delete-laws}
\[
\rreplace(n,u);\rreplace(m,v) \equiv \left\{
\begin{array}{ll}
  \rreplace(m,v) & \text{if $m <_t n$}\\
  \rreplace(m,v);\rreplace(n,u) &\text{if $m \in N_t$, $m \leq_t n$}\\
  \rreplace(n,\semantics{\rreplace(m,v)}{u}) &\text{if $m \in N_u-\{r_u\}$}
\end{array}
\right.
\]
\caption{Chaining and commuting replacements}\label{fig:replace-laws}
\end{figure}

We can use these identities to normalize an update sequence as
follows.  First, move occurrences of inserts to the end of the
sequence.  Next, move deletes to the beginning of the sequence.
Finally, we use the remaining rules to eliminate dependencies among
deletes, replacements and inserts, and to build chains of
replacements.  The resulting sequence is in normal form.

Note that most of the identities only rearrange existing allowed
updates and do not introduce any new update operations that we need to
check against the policy.  In a few cases, we need to do some work to
check that the rewritten sequence is still allowed.  For example, when
we rewrite $\rreplace(n,u);\rdelete(m)$ to $\rdelete(n)$ with $m =
r_u$, we need to verify that we are allowed to delete $m$; this is
because we were allowed to delete $n$, which replaced $m$.
\end{proof}

We say that two trees \emph{agree above $n$} if the trees are equal
after deleting the subtree rooted at $n$ from each.  Note that for all
of the operations we consider, if $op$ has principal node $n$ and $op$
is valid on $t$ then $t$ agrees with $\semantics{op}{t}$ above $n$.

\begin{lemma}
  If $t$ and $t'$ are equal except under the subtree starting at $n$,
  and allowed sequence $\bar{a}$ maps $t$ to $t'$, then there is an
  equivalent, normalized, allowed sequence $\bar{a'}$ that only
  affects nodes at or above $n$.
\end{lemma}
\begin{proof}
  We show that for each node $m$ unrelated to $n$, updates applying
  directly to $m$ can be eliminated.  If a deletion applies to $m$,
  then must be an insertion replacing the deleted subtree exactly, and
  these are the only updates affecting $m$.  Thus, it is safe to
  remove this useless deletion-insertion pair.  If a replacement
  applies to $m$, then there must be subsequent replacements that
  restore the subtree at $m$.  This sequence of replacements can be
  eliminated.  No other possibilities are consistent with $t$ and $t'$
  being equal except at $n$.  Thus, by considering each node $m$ in
  the tree that is unrelated to $n$, and removing the updates having
  an effect on $m$, we can obtain an equivalent update sequence
  $\bar{a'}$ having only updates whose principal node is related to
  $n$.  This update sequence is still allowed since we have only
  removed allowed operations (and since all of the operations we have
  removed are independent of the remaining ones), and can also be
  further normalized if necessary.
\end{proof}

If $t,t'$ agree above $n$, and $\bar{a}$ is an allowed sequence, then
we define the $n$-related normal form of $\bar{a}$ to be an
equivalent allowed, normalized sequence of operations affecting the
tree above or below $n$, which must exist by the  above lemma.

\begin{proofOf}{Theorem~\ref{thm:TPconsis}}
  For the forward direction, we prove the contrapositive.  As argued
  in Section~\ref{sec:consistent-policies}, any violations of the above properties
  suffice to show that a policy is inconsistent.

  For the reverse direction, we again prove the
  contrapositive.  Suppose $P$ is inconsistent, and let $t$ be a tree,
  $\bar{a}$ a sequence allowed on $t$, and $d$ denied on $t$ by $P$,
  such that $\semantics{\bar{a}}{t} = \semantics{d}{t}$.  We consider
  the four cases for $d$:
\begin{itemize}
\item $d = \rinsert(n,t)$.  Consider the normal form of the $\bar{a}$
  restricted to the updates related to $n$.  Clearly $\bar{a}$ cannot
  consist only of updates at or below $n$ since an insertion at $n$
  cannot be simulated by a deletion or replacement at $n$ or by any
  operations that only apply below $n$.  If there is a deletion above
  $n$, there must also be an insertion above $n$ that restores the
  extra deleted nodes and also has the effect of $\rinsert(n,t)$.
  Hence there is a violation of rule 1.  Otherwise, if there is a
  replacement above node $n$, then there must be one or more
  replacements restoring the rest of the tree to its previous form and
  inserting $t$, violating rule 3 (since the chain of replacements
  must be allowed by a cycle in some graph $\G_A$)
\item $d = \rdelete(n,t),\rreplace(n,s)$.  Similar to case for
  $\rinsert$, since again these operations cannot be simulated solely
  by operations at or below $n$.
\item $d = \rreplace(n,v)$.  There are two possibilities.  If the
  $n$-related normal form of $\bar{a}$ consists only of replacements
  at $n$, then the policy must violate rule 2.  Otherwise, an
  argument similar to that in the above cases can be used to show that
  $P$ must violate rule 1 or 3.
\end{itemize}
\vspace{-3mm}
\end{proofOf}

\begin{proofOf}{Proposition~\ref{prop:consis}}
By Theorem~\ref{thm:TPconsis}, there are two cases in which a policy
can be inconsistent. The first case can be checked by doing a
traversing of the graph following a topological sorting of the DTD
graph. This can be done in polynomial time over the number of edges
and vertices of the DTD graph.

The second case consists of checking if the graphs $\G_A$ are acyclic
and transitive. Checking this two conditions for each element $A$ can
be done in polynomial time.
\end{proofOf}

\begin{proofOf}{Lemma~\ref{lem:conj-ext}}
  Since both $P$ and $Q$ extend $R$, we have $\Allow_P,\Allow_Q \supseteq \Allow_R$ and $\Deny_P,\Deny_Q \supseteq \Deny_R$; hence
\begin{eqnarray*}
\Allow_{P \wedge Q} &=& \Allow_P \cap \Allow_Q \supseteq \Allow_R \cap \Allow_R = \Allow_R\\
  \Deny_{P \wedge Q} &=& \Deny_P \cup \Deny_Q \supseteq \Deny_R \cup \Deny_R = \Deny_R
\end{eqnarray*}
\end{proofOf}

\begin{proofOf}{Lemma~\ref{lem:t-simulation}}
  By cases according to the definition of $T$.  If $uat \in S$ then
  there is nothing to do.

If for some $A,$ $B$ we have $uat = (C,op)$ with $B$ $\leq_D$ $C,$ with production rule
  $A$ $\ra$ $B^*,$ $\{(A,$ $\rinsert(B)),$ $(A,$ $\rdelete(B))\}$ $\subseteq$ $S$,
  then let $n = node(op_0)$, let $m$ be the $B$-labeled node above $m$
  in $t$ (there must be exactly one), and let $t'$ be the subtree of
  $t$ rooted at $m$.  We can simulate $op_0$ by deleting the
  $B$-labeled subtree to which $op_0$ applies, then inserting the tree
  resulting from applying $op_0$; thus, the sequence $\bar{op} =
  \rdelete(m); \rinsert(n,\semantics{op_0}{t'}$ simulates $op_0$ and
  is allowed.

  If for some $A,B$ we have $uat = (C,op)$ with $B_i \leq_D C, Rg_D(A) = B_1+\ldots+B_n,
  (B_i,B_i) \in \G_A^+(S)$, then let $B_{i_1},\ldots,B_{i_k}$ be a
  cycle in $\G_A$ beginning and ending with $B_i$.  Again let $n=
  node(op_0)$, $m $ be the (unique) $B_i$-labeled node above $n$, and
  $t'$ be the subtree of $t$ rooted at $m$.  Let $t_1,\ldots,t_{k-1}$
  be arbitrary trees disjoint from $t$ and satisfying $t_j \in
  I_D(B_{i_j})$.  (The latter sets are always nonempty so such trees
  may be found.)  Now consider the update sequence
  \[\small
\begin{array}{c}
\bar{op} =
  \rreplace(m,t_1);\rreplace(rt_{t_1},t_2);\ldots;\rreplace(rt_{t_{n-2}},t_{n-1});\rreplace(rt_{t_n-1},\semantics{op_0}{t'})
\end{array}
\]
  This update sequence is allowed on $t$ and simulates $op_0$.

  Finally, if for some $B_1,\ldots,B_n$ we have $uat =
  (C,\rreplace(B_i,B_j))$, where $Rg_D(C) = B_1+\ldots+B_n, (B_i,B_j) \in
  \G_C^+(S)$ then let $n = node(op_0)$, let $t' $ be the subtree
  rooted at $n$.  Let $B_{i_1},\ldots,B_{i_k}$ be a sequence of nodes
  forming a path from $B_i = B_{i_1}$ to $B_j = B_{i_k}$ in $\G_C$,
  and choose $t_1,\ldots,t_{k-1}$ satisfying $t_l \in I_D(B_{i_l})$.
  Then the update sequence
  \[\small
  \begin{array}{c}\bar{op} =
    \rreplace(n,t_1);\rreplace(rt_{t_1},t_2);\ldots;\rreplace(rt_{t_{n-2}},t_{n-1});\rreplace(rt_{t_n-1},\semantics{op_0}{t'})
  \end{array}
  \]
  again is allowed and simulates $op_0$.
\end{proofOf}

\subsection{Proofs from Section~\ref{sec:rep}}

\begin{proofOf}{Theorem~\ref{thm:RepProblem}} We will concentrate on
the total-repair problem. The proof for partial-repair problem is
analogous.

First we will prove that the total-repair is in \NP. We can determine
if there is a repair $P'=(\A',\F')$ of $P$ such that $|\A \sm \A'| <
k$, by guessing a policy $P'$, checking if $|\A \sm \A'| < k$ and if
it is consistent. Since consistency and the distance can be checked
in polynomial time, the algorithm is in \NP.

To prove that the problem is \NP-hard, we reduce the  edge-deletion
transitive-digraph  problem which is \NP-complete \cite{Yan78,Yan81}.
The problem consists in, given a directed graph $\G=(\V,\E)$ with
$V=\{v_1,\dots, v_n\}$ and $E$ a set of edges without self-loops,
determine if there exists a set $\G'=(\V,\E')$ such that $E'\subseteq
E$, $\G'$ is transitive and $|E \sm E'|<k$. Now, let us define a DTD
$D$ and a policy $P$. The production rules of $D$ are:

$A \ra v_1 + \dots + v_n$\\ \indent$v_i \ra \str$ \hspace{4mm}for $i
\in [1,n]$

\ni The policy $P=(\A,\F)$ is such that $\A=\{(A,\rreplace(v_i,v_j))
| (v_i,v_j) \in E\} \cup \{(v_i,$ $\rreplace(\str,\str)) \mid v_i \in
\V\}$ and $\F=\valid(D) \sm \A$. It is easy to see that $\G_A=\G$ and
therefore finding a repair will consist on finding the minimal number
of edges to delete from $\G$ to make the graph transitive.
\end{proofOf}

\begin{proofOf}{Theorem~\ref{thm:sound}} Given an inconsistency policy
$P=(\A,\F),$ Let us assume, by contradiction, that  the policy
$P'=(\A',\F')$ returned by algorithm $\mathbf{Repair}$ is not a
repair. Since $P'$ is defined over $D$, and by construction $P' \leq
P$, this implies that $P'$ is not consistent. Then, it should be the
case that either the changes returned by:
\begin{enumerate}
  \item {\em $\mathbf{InsDelRepair}$ do not solve all the
      insert/delete-inconsistencies}. This implies that there is
      a node $A$ with production rule $A \ra B*$ such that
      $(A,\rinsert(B))$ $\in$ $\A'$, $(A,\rdelete(B))$ $ \in$
      $\A'$ and there is at least one forbidden \UAT, say
      $(C,op)$, such that $B$ $ \leq_D$ $C$. Since $P' \leq P$,
      $(A,\rinsert(B)) \in \A$ and $(A,\rdelete(B)) \in \A$. If
      we prove that there is always an operation $(G,op) \in \F$
      such that $B$ $ \leq_D$ $G$, the marked DTD graph would be
      such that $\chi(A)=\perp$. Then, either $(A,\rinsert(B))$
      or $(A,\rdelete(B))$ would have been in the changes
      returned by $\mathbf{InsDelRepair}$ and one of them
      wouldn't have belonged to $P'$. Now we will prove that such
      $(G,op)$ always exists. If $(C,op) \in \F$, then,
      $(G,op)=(C,op)$. On the other hand, if $(C,op) \not \in \F$
      then $(C,op)$ is either one of the changes
returned by $\mathbf{InsDelRepair}$ or $\mathbf{ReplaceRepair}$:
\begin{enumerate}
     \item If $(C,op)$ was a change returned by
         $\mathbf{InsDelRepair}$, then there was an
         insert-delete inconsistency, and there is another
         \UAT $(F,op2) \in \F$ such that $C$ $ \leq_D$ $F$.
         As a consequence $B$ $ \leq_D$ $F$, and we have
         found $(G,op)$.
         \item If $(C,op)$ was a change returned by
             $\mathbf{ReplaceRepair}$ this would mean that
             $(C,op)$ was either involved in a negative-cycle
             or forbidden-transitivity. The former implies
             there is another \UAT $(F,op2) \in \F$ such that
             $C$ $ \leq_D$ $F$. Then, $B$ $ \leq_D$ $F$, and
             we have found $(G,op)$. The latter case implies
             there is at least one other $(C,op2)\in F$. We
             have found $(G,op)$.
\end{enumerate}

  \item {\em $\mathbf{ReplaceRepair}$ do not solve all the
      replace-inconsistencies}: This implies that there is a node
      $A$ with production rule $A \ra B_1+ \dots + B_n$ such that
      one of the following holds:
\begin{enumerate}

 \item There is an edge $(B_i,B_j)$ in $\G_A^+$ for $P'$,
     s.t. $(B_i,B_j) \in \F'_A$.  If $(B_i,B_j) \in \F_A$,
     then $\mathbf{ReplaceRepair}$ would have deleted at
     least one edge from each justification of $(B_i,B_j)$,
     and therefore, $(B_i,B_j)$ could not be in $\G_A^+$ for
     $P'$. On the other hand, if $(B_i,B_j) \not \in \F_A$,
     then $(A,\rreplace(B_i,B_j))$ it implies that it was
     part of the changes returned by
     $\mathbf{ReplaceRepair}$. Since both,
     $\mathbf{ReplaceNaive}$ and $\mathbf{ReplaceSetCover}$
     check that the final graph has no
     forbidden-transitivity, this is not possible.

\item  There is a $B_i$ which is part of a  cycle in $\G_A$
    for $P'$ and there is a \UAT $(C,op) \in \F'$ s.t. $B_i
    \leq_D C$. Since $B_i$ is in a cycle in $\G_A$ for $P'$,
    it should be part of a cycle in $\G_A$ for $P$. If
    $(C,op) \in \F$, then the inconsistency would have been
    solve. On the other hand, if $(C,op) \not \in \F$, then
$(C,op)$ is
      either one of the changes returned by
$\mathbf{InsDelRepair}$ or $\mathbf{ReplaceRepair}$. By an
analogous reasoning as in cases 1(a)-1(b), this is not
possible either.
\end{enumerate}
\end{enumerate}
Therefore, $P'$ is consistent and is a repair of $P$.
\end{proofOf}

\section{Algorithms}
\label{app:alg}

\begin{algorithm}[ht]
\caption{$\mathbf{markGraph}$ } \label{alg:markgraph}
\begin{algorithmic}[1]
 \REQUIRE DTD Graph $\sdg{D}$, Policy $P$
 \ENSURE Marked DTD Graph $\mg{D} = (\sdg{D}, \mu, \chi)$
 \STATE Let $l_1, l_2, \ldots l_k$ be the set of nodes in $\sdg{D}$ with out-degree=$0$
 \FORALL{$l$ in $\{l_1, l_2, \ldots l_k \}$}
 \STATE $\mathbf{markNode}(\mg{D},l, P)$
 \ENDFOR
  \RETURN $\mg{D}$
\end{algorithmic}
\end{algorithm}

\begin{algorithm}[ht]
\caption{$\mathbf{markNode}$ }
\label{alg:marknode}
\begin{algorithmic}[1]
 \REQUIRE Marked DTD Graph $\mg{D} = (\sdg{D}, \mu, \chi)$, Node $B$, Policy $P = ({\cal A}, {\cal F})$
 \FORALL{$A \in \V_{\dg{D}}$ such that $(A,B) \in E_{\dg{D}}$}
 \IF{$\mu(B) = \mathquote{-}$}
 \STATE $\mu(A) \la \mathquote{-}$
 \ELSE
 \STATE \algcom{$\mu(B)$ is undefined}
 \IF{$(A, \rinsert(B)) \in \F$ or $(A, \rdelete(B))\in \F$ or $(A, \rreplace(B, B'))\in \F$ }
 \STATE $\mu(B)\la \mathquote{-}$, $\mu(A) \la \mathquote{-}$
 \ELSE
 \STATE $\mu(B) = \mathquote{+}$
 \ENDIF
 \ENDIF
\IF {$\mu(A) = \mathquote{-}$}
 \IF{$(A, \rinsert(B)) \in \A$ and $(A, \rdelete(B))\in \A$}
 \STATE $\chi(A) \la \mathquote{\perp}$
 \ENDIF
\ENDIF
 \STATE $\mathbf{markNode}(A)$
 \ENDFOR
 \end{algorithmic}
 \end{algorithm}

\begin{algorithm}[ht]
\caption{$\mathbf{InsDelRepair}$ }
\label{alg:findinconsistencies}
\begin{algorithmic}[1]
 \REQUIRE DTD graph $\sdg{D}$, security policy $P$
 \ENSURE Set of \UATs to remove from $P$ to restore consistency in $P$ \wrt
 insert/delete-inconsistencies
  \STATE $\mg{D} \la \mathbf{markGraph}(\sdg{D}, P)$
  \STATE $\mathit{changes} \la \emptyset$
  \FORALL {$A \in \V_{\dg{D}}$ and $(A,B) \in E_{\dg{D}}$ }
  \IF {$\chi(A) = \mathquote{\perp}$}
   \STATE Randomly choose either $(A, \rinsert(B)$ or $(A,
   \rdelete(B))$ and assign it to $U$
   \STATE $\mathit{changes} \la \mathit{changes} \cup U$
   \ENDIF
   \ENDFOR
  \RETURN $\mathit{changes}$
\end{algorithmic}
\end{algorithm}

\begin{algorithm}[ht]
\caption{$\mathbf{ReplaceRepair}$ } \label{alg:replace_checking}
\begin{algorithmic}[1]
 \REQUIRE DTD graph $\sdg{D}$, security policy $P=(\A,\F)$, Maximum Number of Justifications $\MNJ$
\ENSURE Set of \UATs to remove from $\A$ to restore consistency in $P$ \wrt
 replace-inconsistencies
 \STATE $\mg{D} \la \mathbf{markGraph}(\sdg{D},P)$
 \IF{$\MNJ = 0$}
 \STATE $\mathit{Sol} \la \mathbf{ReplaceNaive}(r_{\dg{D}}, \mg{D})$
 \ELSE
 \STATE $\mathit{Sol} \la \mathbf{ReplaceSetCover}(r_{\dg{D}}, \mg{D}, \MNJ)$
 \ENDIF
 \STATE $\mathit{changes} \la \emptyset$
 \FORALL{$(A,\C) \in \mathit{Sol}$}
    \FORALL{ $(B,C)\in \C$ }
        \STATE $\mathit{changes} \la \mathit{changes} \cup (A,\rreplace(B,C))$
    \ENDFOR
 \ENDFOR

 \RETURN $\mathit{changes}$
\end{algorithmic}
\end{algorithm}

\begin{algorithm}[ht]
\caption{$\mathbf{ReplaceNaive}$ }
\label{alg:replace_naive}
\begin{algorithmic}[1]
 \REQUIRE Node $R$, Marked Graph $\mg{D}$
 \ENSURE Set $Sol$ containing pairs $(B,\C)$ where $B$ is a node
reachable from $R$ in $\mg{D}$, and $\C$ a  set of edges to delete
from $\G_B$ to make it consistent
 \IF {$Rg(R):=B_1+B_2\ldots+B_n$}
 \STATE Let $\G_R$ be the replace graph for $R$
 \STATE $\C \la \emptyset$
 \STATE Let stack $S$ contain all the nodes in  c
 \WHILE{$S$ not empty}
  \STATE $B \la S.pop()$
 \FORALL{$A$ in $V_R$, s.t. $(A,B) \in \E_R \setminus \C$}
 \FORALL{$C \in V_R$, s.t. $(B,C) \in \E_R \setminus \C$}
 \STATE \algcom{If there is an edge missing for transitive or if there is a cycle over a node with a \UAT forbidden below}
 \IF {$A \not = C$ or $\mu(A) = \mathquote{-}$}
    \STATE Let $e$ be one of $(A,B)$, $(B,C)$ (chosen randomly)
    \STATE $\C = \C \cup \{ e \}$
    \IF {$e=(A,B)$}
        \STATE $G=A$
    \ELSE
        \STATE $G=B$
    \ENDIF
    \FORALL {$F \in \V_R$ s.t. $F$ is reachable from $G$ in  $\G_R$}
        \STATE $S.push(F)$
    \ENDFOR
 \ENDIF
 \ENDFOR
 \ENDFOR
 \ENDWHILE
 \STATE $Sol \la \{(R,\C)\}$
 \ELSE
 \STATE $Sol \la \emptyset$
 \ENDIF
 \FORALL{$(R,B) \in \E_R$}
 \STATE $Sol \la Sol \cup \mathbf{ReplaceNaive}(B, \mg{D})$
 \ENDFOR
 \RETURN $Sol$
\end{algorithmic}
\end{algorithm}

\begin{algorithm}[ht]
\caption{$\mathbf{ReplaceSetCover}$ } \label{alg:replace_traverse}
\begin{algorithmic}[1]
 \REQUIRE Node $R$, marked DTD graph $\mg{D}$,  forbidden edges $\F_R$, integer $\MNJ$
 \ENSURE Set $Sol$ containing pairs $(B,\C)$ where $B$ is a node
reachable from $R$ in $\mg{D}$, and $\C$ a set of edges to delete
from $\G_B$ to make it consistent
 \STATE $Sol \la \emptyset$,~ $\C \la \emptyset$,~  {\em done$\la$ false}
 \IF{$Rg(R):=B_1+B_2\ldots+B_n$}
    \STATE Let $\G_R=(\V,\E)$ be the replace graph for $R$
    \STATE $\G \la \G_R$
    \WHILE {$\neg${\em done}}
         \STATE $\G^+ \la \mathbf{ComputeJustifications}(\G, \MNJ)$
         \STATE \algcom{Algorithm $\mathbf{setCoverAlg}$ takes the graph $\G^+$
with the justifications and the set of forbidden edges and returns
the edges to delete from $\G_A$}
         \STATE $\sce \la \mathbf{setCoverAlg}(\G^+,\F_R)$
          \IF {$\sce \not = \emptyset$}
            \STATE remove edges in $\sce$ from $\G$
            \STATE $\C \la \C \cup \sce$
          \ELSE
            \STATE {\em done = true}
          \ENDIF
    \ENDWHILE
        \STATE $Sol \la Sol \cup \{(R,\C)\}$
 \ENDIF
 \FORALL{$(R,B) \in \E_{R}$}
 \STATE $Sol \la Sol \cup \mathbf{ReplaceSetCover}(B, \mg{D})$
 \ENDFOR
\RETURN $Sol$
\end{algorithmic}
\end{algorithm}

\begin{algorithm}[ht]
\caption{ $\mathbf{ComputeJustifications}$ } \label{alg:traverse2}
\begin{algorithmic}[1]
 \REQUIRE Replace Graph $\G_{R}$,  Maximum Number of Justifications $\MNJ$
 \ENSURE $\G_{R}^{+}$, \ie  the transitive closure of $\G_{R}$
with each edge and node labelled with a set $\J$ containing at most
$\MNJ$ justifications
 \STATE $E \la \emptyset$
 \FORALL{$(A,B) \in \E_{R}$}
    \STATE $\J((A,B)) = \{ \{ (A, B) \} \}$
 \ENDFOR
 \FORALL{$A \in \V_{R}$}
    \STATE $\J(A) = \emptyset$
 \ENDFOR
 \FORALL{$A$ in $\V_{R}$}
    \FORALL{$B$ in $\V_{R}$, s.t. $(A,B) \in \E_{R} \cup E$}
        \FORALL{$C \in V_{R}$, s.t. $(B,C) \in \E_{R} \cup E$}
            \STATE \algcom{If there is an edge missing for transitivity}
            \IF{$(A, C) \not \in \E_{R}$ and $A \not = C$}
                \IF{$(A,C) \not \in E$}
                    \STATE $E \la E \cup \{ (A, C) \}$
                    \STATE $\J((A,C)) \la \emptyset$
                \ENDIF
                \FORALL {$j_1 \in \J((A,B))$}
                    \FORALL {$j_2 \in \J((B,C))$}
                        \IF{ $|\J((A,C))|$ $<$ $\MNJ$}
                            \STATE $\J((A,C)) \la \J((A,C)) \cup \{ j_1 \cup j_2 \}$
                        \ENDIF
                    \ENDFOR
                \ENDFOR
            \ENDIF
            \STATE \algcom{If there is a cycle}
            \IF{$A=C$ and $\mu(A) = \mathquote{-}$}
                \FORALL {$j_1 \in \J((A,B))$}
                    \FORALL {$j_2 \in \J((B,A))$}
                        \IF{ $|\J(A)|$ $<$ $\MNJ$}
                            \STATE $\J(A) \la \J(A) \cup \{ j_1 \cup j_2 \}$
                        \ENDIF
                    \ENDFOR
                \ENDFOR
            \ENDIF
        \ENDFOR
    \ENDFOR
 \ENDFOR
 \STATE $\G_{R}^{+} \la (\V_{R}, \E_{R} \cup E)$
 \RETURN $\G_{R}^{+}$
\end{algorithmic}
\end{algorithm}

\begin{algorithm}[ht]
\caption{$\mathbf{Repair}$} \label{alg:repair}
\begin{algorithmic}[1]
 \REQUIRE DTD graph $G_D$, security policy $P = ({\cal A},{\cal F})$,
 boolean {\em total}
 \ENSURE A repair $P'$ of $P$. The repair is total if parameter {\em
 total}$=1$, partial otherwise.
 \STATE $\mathit{changes} \la \mathbf{InsDelChecking}(G_D,P) \cup \mathbf{ReplaceRepair}(G_D,P)$
 \STATE  $\A' \la \A - \mathit{changes}$
  \IF {{\em total}}
  \STATE $\F' \la \valid(D) - \A'$
  \ELSE
  \STATE $\F' \la \F$
  \ENDIF
  \STATE $P'\la (\A',\F')$
 \RETURN $P'$
\end{algorithmic}
\end{algorithm}

\begin{figure}
\vspace{-2ex}
\begin{center}
\fbox{
\includegraphics[scale=0.45]{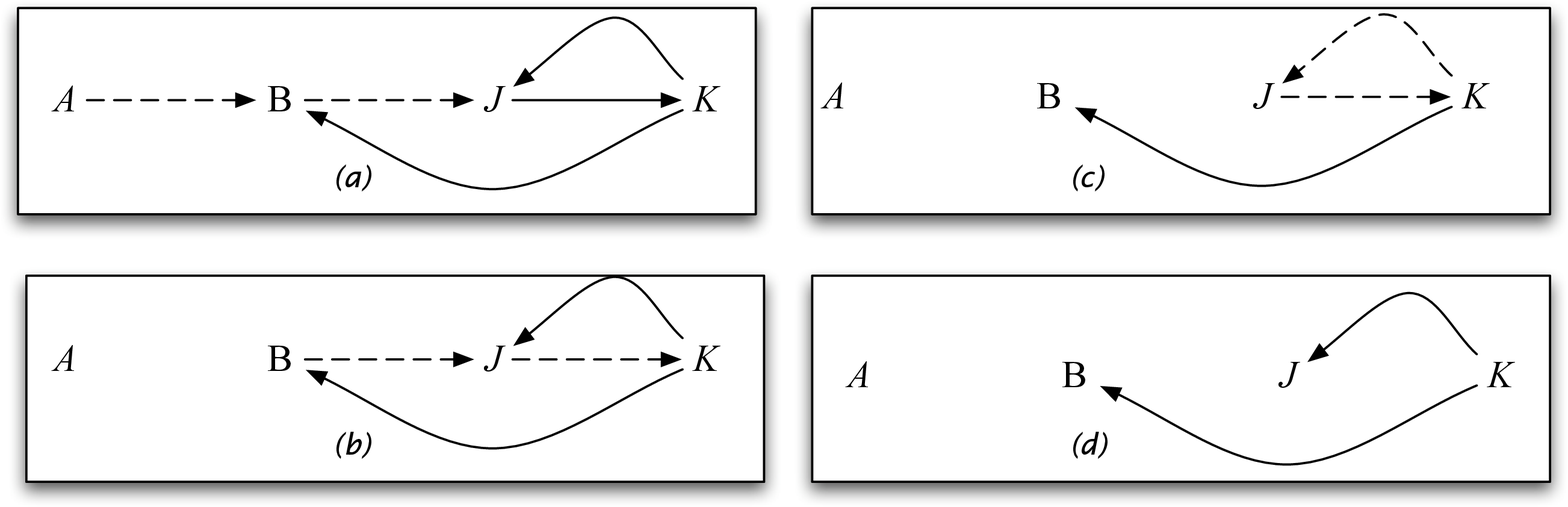}
}
\end{center}\vspace{-2ex}
\caption{Execution of $\naive$ on $\G_{R}$}
\label{fig:rreplace-naive}
\vspace{-2ex}\end{figure}



\end{document}